\documentclass[12pt]{article}
\pdfoutput=1

\usepackage{amsmath}
\usepackage{amsfonts}
\usepackage{amscd}
\usepackage{amsthm}
\usepackage{setspace}

\usepackage{graphicx}
\usepackage{authblk}
\usepackage{caption}
\usepackage{ytableau}
\usepackage{mathtools}

\usepackage[OT2,T1]{fontenc}
\DeclareSymbolFont{cyrletters}{OT2}{wncyr}{m}{n}
\DeclareMathSymbol{\Sha}{\mathalpha}{cyrletters}{"58}

\def\Z{\mathbb{Z}}

\def\P{\mathbb{P}}

\begin{document}

\begin{titlepage}

\begin{flushright}
KEK-TH-2148
\end{flushright}

\vskip 1cm

\begin{center}

{\bf \Large F-theory models with $U(1)\times \mathbb{Z}_2,\, \mathbb{Z}_4$\\
\vspace{0.5cm}
and transitions in discrete gauge groups}

\vskip 1.2cm

Yusuke Kimura$^1$ 
\vskip 0.6cm
{\it $^1$KEK Theory Center, Institute of Particle and Nuclear Studies, KEK, \\ 1-1 Oho, Tsukuba, Ibaraki 305-0801, Japan}
\vskip 0.4cm
E-mail: kimurayu@post.kek.jp

\vskip 2cm
\abstract{We examine the proposal in the previous paper to resolve the puzzle in transitions in discrete gauge groups. We focus on a four-section geometry to test the proposal. We observed that a discrete $\Z_2$ gauge group enlarges and $U(1)$ also forms in F-theory along any bisection geometries locus in the four-section geometry built as the complete intersections of two quadrics in $\P^3$ fibered over any base. Furthermore, we demonstrate that giving vacuum expectation values to hypermultiplets breaks the enlarged $U(1)\times \Z_2$ gauge group down to a discrete $\Z_4$ gauge group via Higgsing. We thus confirmed that the proposal in the previous paper is consistent when a four-section splits into a pair of bisections in the four-section geometry. This analysis may be useful for understanding the Higgsing processes occurring in the transitions in discrete gauge groups in six-dimensional F-theory models. We also discuss the construction of a family of six-dimensional F-theory models in which $U(1)\times\Z_4$ forms.}  

\end{center}
\end{titlepage}

\tableofcontents
\section{Introduction}
F-theory \cite{Vaf, MV1, MV2} is compactified on manifolds that admit a torus fibration. Axiodilaton in type IIB superstrings and the modular parameter of elliptic curves as fibers of the torus fibration are identified in F-theory, enabling the axiodilaton to exhibit $SL_2(\Z)$ monodromy. 
\par In recent years, F-theory compactifications on genus-one fibrations without a global section have attracted interest, for reasons including the discrete gauge group \footnote{See, e.g., \cite{KNPRR, ACKO, BS, HSsums, CIM, BISU, ISU, BCMRU, BCMU, MRV, HS, BRU, KKLM, HS2, GPR} for recent progress of discrete gauge groups.} arising in this type of compactification \cite{MTsection} of F-theory. 
\par There are situations in which a genus-one fibration has a global section and in which it does not have a global section; when a genus-one fibration does not have a global section, a discrete gauge group forms in F-theory on this fibration, as mentioned. Recent discussions of F-theory on genus-one fibrations without a global section can be found, for example, in \cite{BM, MTsection, AGGK, KMOPR, GGK, MPTW, MPTW2, BGKintfiber, CDKPP, LMTW, K, K2, ORS1604, KCY4, CGP, Kdisc, Kimura1801, AGGO1801, Kimura1806, TasilectWeigand, CLLO, TasilectCL, HT, Kimura1810, Kimura1902, Kimura1905, Kimura1907} \footnote{\cite{BEFNQ, BDHKMMS} discussed F-theory on genus-one fibrations without a global section.}. When a genus-one fibration has a global section (in which case, the fibration is often called an elliptic fibration \footnote{See, e.g. \cite{MorrisonPark, MPW, BGK, BMPWsection, CKP, BGK1306, CGKP, CKP1307, CKPS, Mizoguchi1403, AL, EKY1410, LSW, CKPT, CGKPS, MP2, MPT1610, BMW2017, CL2017, BMW1706, KimuraMizoguchi, Kimura1802, LRW2018, MizTani2018, CMPV1811, TT2019, Kimura1903, EJ1905, LW1905} for discussions of F-theory on elliptic fibrations with a global section.} in the F-theory literature), the $U(1)$ gauge group forms in F-theory if the fibration has two or more independent global sections. The global sections of an elliptic fibration form a group, known as the Mordell--Weil group, which has the notion of ``rank.'' The rank of the Mordell--Weil group is given by one less the number of independent global sections. The rank of the Mordell--Weil group gives the number of $U(1)$s arising in F-theory on the elliptic fibration \cite{MV2}. 

\vspace{5mm}

\par A genus-one fibration lacking a global section still has a ``multisection,'' and whereas a global section (which can be seen as a ``horizontal'' divisor) intersects a fiber (which can be seen as a ``vertical'' divisor) at one point, a multisection of ``degree $n$,'' or more concisely, an ``$n$-section,'' intersects a fiber in $n$ points. A discrete $\Z_n$ gauge group forms in F-theory on a genus-one fibration with an $n$-section \cite{MTsection} \footnote{When F-theory is compactified on a Calabi--Yau genus-one fibration $Y$, the discrete gauge group arising in this compactification is identified with the discrete part of the ``Tate--Shafarevich group,'' $\Sha(J(Y))$, of the Jacobian, $J(Y)$ \cite{BDHKMMS, MTsection}.}. 
\par In the moduli of the $n$-section, an $n$-section deforms to split into $n$ sheets of separate global sections. It was argued in \cite{MTsection} that the physical viewpoint of this process reverses the geometric order and can be interpreted as a Higgsing process wherein $U(1)^{n-1}$ breaks down into a discrete $\Z_n$ gauge group. 

\vspace{5mm}

\par However, there are various other manners in which a multisection splits into multisections of smaller degrees in the moduli of multisection geometry. When studying these, physically unnatural phenomena are identified \cite{Kimura1905, Kimura1907}. Under certain conditions, a four-section splits into a pair of bisections \cite{Kdisc, Kimura1905}. When one considers the process \cite{Kimura1905} wherein a four-section splits into a pair of bisections, and these bisections further split into four global sections, when seen from the physical viewpoint, $U(1)^3$ breaks into a discrete $\Z_2$ gauge group, and this discrete $\Z_2$ gauge group transitions further to a discrete $\Z_4$ gauge group via Higgsing. It was pointed out in \cite{Kimura1905} that this process appears unnatural because a discrete $\Z_2$ gauge group appears ``enhanced'' to a discrete $\Z_4$ gauge group, rather than broken down into another discrete gauge group of a smaller degree. In contrast, another puzzling physical phenomenon was observed in \cite{Kimura1907}, wherein an $n$-section (with $n\ge 3$) splits into a global section and an $(n-1)$-section. This process can be viewed from the physical viewpoint as a Higgsing process wherein an F-theory model without a discrete or $U(1)$ gauge group transitions to another model with a discrete $\Z_n$ gauge group, and this also appears puzzling \cite{Kimura1907}. 
\par It was proposed in \cite{Kimura1907} that if one interprets a gauge group to be enhanced to a larger gauge symmetry at the points in the multisection geometry where a multisection splits into multisections of smaller degrees, these apparently puzzling phenomena can be naturally explained. This proposal includes a situation in which a discrete $\Z_2$ gauge group is enhanced to $U(1)\times \Z_2$ \cite{Kimura1907} \footnote{See also, e.g., \cite{KMOPR, GKK, CLLO} for discussions of F-theory models in which both $U(1)$ and a discrete gauge symmetry form.}. 

\vspace{5mm}

\par The aim of this note is to examine whether the proposal in \cite{Kimura1907} to resolve the puzzle pointed out in \cite{Kimura1905} is consistent along bisection geometry loci in a four-section geometry. We focus on six-dimensional F-theory compactifications to test the proposal. We confirm in Section \ref{section3.1} that a discrete $\Z_2$ gauge group actually expands and $U(1)$ also forms in F-theory along any bisection geometry locus in a four-section geometry, realized as the complete intersections of two quadrics in $\P^3$ fibered over any base space. Furthermore, we demonstrate that the Higgsing process giving vacuum expectation values (vevs) to hypermultiplets breaks the enlarged $U(1) \times \Z_2$ group down to a discrete $\Z_4$ gauge group, thus confirming that the proposal given in \cite{Kimura1907} indeed occurs along the bisection geometry loci in the four-section geometry, when the base spaces are isomorphic to $\P^1\times \P^1$ and $\P^2$. These observations can support, at least to some degree, the proposal in \cite{Kimura1907}.
\par Points in the moduli of multisection geometry on which both $U(1)$ and discrete gauge groups form simultaneously are relevant to the physically puzzling phenomena noted in \cite{Kimura1905, Kimura1907}, and studying models in which both $U(1)$ and discrete gauge groups form can be useful in checking the consistency of the proposal in \cite{Kimura1907} to resolve the puzzles. Motivated by this background, we also construct a family of F-theory models in which $U(1) \times \Z_4$ forms in Section \ref{section4}. 

\vspace{5mm}

\par The number of $U(1)$ arising in F-theory on a genus-one fibration lacking a global section is given by the Mordell--Weil rank of its Jacobian fibration \footnote{The relation of the moduli of genus-one fibrations and those of the Jacobian fibrations and Weierstrass models was discussed in \cite{MTsection} in the context of F-theory.} \footnote{Construction of the Jacobians of elliptic curves is discussed in \cite{Cas}.} \cite{BM}, whose types of singular fibers and the discriminant locus are identical to those of the original genus-one fibration.  

\vspace{5mm}

\par We consider in this study Calabi--Yau four-section geometry built as complete intersections of two quadrics in $\P^3$ \cite{BGKintfiber, Kdisc, Kimura1905} fibered over any base. This approach has several advantages. One of the advantages of this approach is that the four-section geometry realized in this fashion contains a bisection geometries locus \cite{Kdisc, Kimura1905, Kimura1907}. Another advantage is that one can construct the associated double cover of four-section geometry realized as a complete intersection. It can be determined when the Mordell--Weil rank of the Jacobian increases to one by studying the coefficients of this associated double cover \cite{MTsection}.

\vspace{5mm}

\par Recent model buildings of F-theory emphasized the use of local model buildings \cite{DWmodel, BHV1, BHV2, DW}. Global aspects of models, however, need to be studied to address the issues of early universe including inflation and the issues of gravity. The compactification geometries are analyzed from the global perspective here.

\vspace{5mm}

\par We present a summary of the results obtained in this work in Section \ref{section2}. We show in Section \ref{section3.1} that a discrete $\Z_2$ gauge group expands and is enhanced to $U(1)\times \Z_2$ along every bisection geometry locus in the four-section geometry, constructed as complete intersections of two quadrics fibered over any base. In Section \ref{section3.3}, we demonstrate that the expanded $U(1)\times \Z_2$ breaks down to a discrete $\Z_4$ gauge group via Higgsing by giving vevs to hypermultiplets, when the base surfaces are isomorphic to $\P^1\times \P^1$ and $\P^2$. 
\par We briefly discuss an example of the construction of six-dimensional F-theory models, in which the $U(1)\times \Z_4$ gauge group forms, in Section \ref{section4}. We state our concluding remarks and open problems in Section \ref{section5}.

\section{Summary of the discussion}
\label{section2}

As pointed out in \cite{Kimura1905, Kimura1907}, physically puzzling phenomena can be observed when some splitting processes in the multisection geometry are analyzed. It was proposed in \cite{Kimura1907} that an interpretation that a (discrete) gauge group tends to enlarge at such points in the moduli where multisections split into multisections of smaller degrees, which can naturally explain the puzzling phenomena.
\par We demonstrate in Section \ref{section3.1} that along any bisection geometries locus in the four-section geometry realized as complete intersections of two quadratic hypersurfaces in $\P^3$ fibered over any base space, $U(1)\times \Z_2$ forms. This confirms that the proposal \cite{Kimura1907} of the puzzle raised in \cite{Kimura1905} is consistent when a four-section is split into a pair of bisections.  

\vspace{5mm}

\par The geometric aspects of the constructions we describe in Sections \ref{section3.1}, \ref{section3.2} and \ref{section4} do not depend on the dimension of the space. Our argument particularly applies both to six-dimensional (6D) and four-dimensional (4D) F-theory models, at least at the geometrical level. However, when one considers four-dimensional F-theory models, the issues of flux \cite{BB, SVW, W96, GVW, DRS} \footnote{Recent progress of F-theory compactifications with four-form flux can be found, for example, in \cite{MSSN, CS, MSSN2, BCV, MS, KMW, GH, KMW2, IJMMP, KSN, CGK, BCV2, LMTW, SNW}.}, including the effect of the superpotential that it generates, also need to be considered \cite{MTsection}. To this end, we mainly focus on 6D F-theory models in this note \footnote{Genus-one fibration structures of 3-folds are analyzed in \cite{Nak, DG, G}.} to discuss F-theory models with $U(1)\times \Z_2$ and $U(1)\times \Z_4$ gauge groups in relation to transitions in discrete gauge groups. We do not discuss the effects of flux in this study.

\vspace{5mm}

\par In this study, we consider, as four-section geometry \cite{BGKintfiber, Kdisc, Kimura1905}, the complete intersection of two quadric hypersurfaces in $\P^3$ fibered over any base space. The general form of this type of complete intersection is given by the following equation:
\begin{eqnarray}
\label{general complete intersection in 2.1}
a_1\, x^2+a_2\, y^2+a_3\, z^2+a_4\, w^2 &  \\ \nonumber
+ 2a_5\,xy+2a_6\,xz+2a_7\,xw+2a_8\,yz+2a_9\,yw+2a_{10}\,zw & =0 \\ \nonumber
b_1\, x^2+b_2\, y^2+b_3\, z^2+b_4\, w^2 &  \\ \nonumber
+ 2b_5\,xy+2b_6\,xz+2b_7\,xw+2b_8\,yz+2b_9\,yw+2b_{10}\,zw & = 0 .
\end{eqnarray}
$[x:y:z:w]$ gives the coordinates of $\P^3$, and $a_i$ and $b_j$, $i,j = 1, \ldots, 10$, are sections of line bundles \footnote{These are subject to the conditions, so when the Jacobian fibration is taken, the Weierstrass form of which is $y^2=x^3 + f\, x+g$, then $[f]=-4K$, $[g]=-6K$, to ensure that the total space yields the Calabi--Yau genus-one fibration, as described in \cite{Kimura1905}. ($K$ denotes the canonical divisor of the base space.)} over the base. For these complete intersections, a method to construct the Jacobian fibration is known \cite{BM, Kimura1905}. This method is described in \cite{BM, Kimura1905}. The construction of the Jacobian \footnote{\cite{BGKintfiber} also discussed a method to construct the Jacobian fibrations of genus-one fibrations built as the complete intersections. We take a different approach here.} consists of two steps: one can build the associated double cover of a quartic polynomial from the complete intersection, and this double cover is generally a bisection geometry \cite{BM, MTsection}. The associated double cover can be expressed as follows:
\begin{equation}
\tau^2 = e_0\lambda^4 + e_1\lambda^3 +e_2\lambda^2 + e_3\lambda +e_4.
\end{equation}
The Jacobian fibration of the double cover is known \cite{BM, MTsection}, and the resulting Jacobian of the double cover yields the Jacobian fibration of the original complete intersection. 
\par When either the constant term $e_4$ or the coefficient $e_0$ of $\lambda^4$ of the double cover is a perfect square, the double cover admits two global sections \cite{MTsection}. In this case, the double cover yields the Jacobian of the complete intersection and it has Mordell--Weil rank is one. Therefore, to demonstrate that along a bisection geometries locus in the four section geometry a discrete $\Z_2$ gauge group enlarges and  $U(1)$ also forms, it suffices to compute the associated double cover and confirm that either of the coefficients $e_0$ or $e_4$ is a perfect square.

\vspace{5mm}

\par We demonstrate in Section \ref{section3.1} that the equation of every bisection geometry locus in the four-section geometry (\ref{general complete intersection in 2.1}), built as the complete intersections of two quadrics in $\P^3$ fibered over any base, admits a transformation to a certain form of complete intersection. We then confirm that the Jacobian of this specific form of complete intersection describing a bisection geometry locus has the Mordell--Weil rank one. This shows that $U(1)\times \Z_2$ forms in F-theory along any bisection geometry locus in the four-section geometry (\ref{general complete intersection in 2.1}).

\par We find that, when some of the parameters of the quadrics of the complete intersections are set to zero along the bisection geometry locus, the $U(1)$ gauge group is enhanced to $SU(4)\times SU(2)\times SU(2)\times SU(2)$. This is discussed in Section \ref{section3.2}.

\vspace{5mm}

\par We demonstrate in Section \ref{section3.3} that giving vevs to hypermultiplets breaks $U(1)\times \Z_2$ that forms along the bisection geometry loci in the four-section geometry down to a discrete $\Z_4$ gauge group via Higgsing. 
\par We deduce the matter spectra on F-theory on the bisection geometry locus when the gauge group is enhanced to $SU(4)\times SU(2)\times SU(2)\times SU(2)\times \Z_2$. We focus on the situations where the base surfaces are isomorphic to $\P^1\times \P^1$ and $\P^2$. The curve in the base surface supporting the $SU(4)$ gauge group has positive genus; the genus is 9 when the base is isomorphic to $\P^1\times \P^1$, and the genus of the curve is 10 when the base is isomorphic to $\P^2$. From these facts, it follows that the adjoint \footnote{\cite{MTsection} discussed adjoint hypermultiplets of $SU(2)$ on the curves of positive genus supporting $SU(2)$ gauge group in 6D F-theory with $SU(2)\times SU(2)$ gauge group on bisection geometries.} hypermultiplets ${\bf 15}$ arise on the curve supporting $SU(4)$. We also determine the matter fields localized at the intersections of the curves supporting $SU(2)$ factors and the curve supporting $SU(4)$. The deduced matter spectra satisfy the 6D anomaly cancellation conditions \cite{GSW6d, Sagnotti, Erler, Sch6d, GM0005, KMT1008}. Giving vevs to hypermultiplets breaks $SU(4)\times \Z_2$ to $U(1)\times \Z_2$. The remaining adjoint hypermultiplets ${\bf 15}$ become scalar fields with $U(1)$ charge 4 through this Higgsing process. Giving vev to one of these scalar fields breaks the $U(1)$ gauge group down to a discrete $\Z_4$ gauge group. This mechanism explains how the $U(1)\times \Z_2$ gauge group breaks down to a discrete $\Z_4$ gauge group via Higgsing.

\vspace{5mm}

\par As noted in the introduction, this result can support to some degree the interpretation proposed in \cite{Kimura1907} to possibly resolve the puzzle in \cite{Kimura1905}, when a four-section is split into a pair of bisections. 

\vspace{5mm}

\par We provide in Section \ref{section4} an example of a family on which $U(1)\times \Z_4$ forms in F-theory. Models with $U(1)\times \Z_4$ can be relevant to the situation where a multisection splits into multisections including a four-section.

\section{Six-dimensional F-theory models with $U(1)\times \Z_2$ and transitions to $\Z_4$ through Higgsing}
\label{section3}

\subsection{Expansion of a discrete gauge group in bisection geometry loci in four-section geometry}
\label{section3.1}
Complete intersection of two quadrics in $\P^3$ fibered over a base space:
\begin{eqnarray}
\label{general complete intersection in 3.1}
a_1\, x^2+a_2\, y^2+a_3\, z^2+a_4\, w^2 &  \\ \nonumber
+ 2a_5\,xy+2a_6\,xz+2a_7\,xw+2a_8\,yz+2a_9\,yw+2a_{10}\,zw & =0 \\ \nonumber
b_1\, x^2+b_2\, y^2+b_3\, z^2+b_4\, w^2 &  \\ \nonumber
+ 2b_5\,xy+2b_6\,xz+2b_7\,xw+2b_8\,yz+2b_9\,yw+2b_{10}\,zw & = 0
\end{eqnarray}
yields a four-section geometry \cite{BGKintfiber, Kimura1905}. $a_i, b_j$ are sections of line bundles, subject to certain conditions, so the genus-one fibration yields a Calabi--Yau manifold \cite{Kimura1905} as previously noted in Section \ref{section2}. $[x:y:z:w]$ are the coordinates of $\P^3$.
\par We now show that when a four-section in the complete intersection splits into a pair of bisections, the equation (\ref{general complete intersection in 3.1}) reduces to a specific form. A four-section splits into bisections, precisely when one of the two quadrics in the complete intersection (\ref{general complete intersection in 3.1}) splits into linear factors along the vanishing of a certain linear equation. After some change of coordinate variables, one can assume that one of the two quadrics in the complete intersection (\ref{general complete intersection in 3.1}) reduces to $xy$ along the vanishing of a certain linear equation. Therefore, a four-section splits into bisections when one of the two quadrics in the complete intersection (\ref{general complete intersection in 3.1}) takes the form: $\alpha\, xy+ (ax+by+cz+dw)(ex+fy+gz+hw)=0$. Under a change of coordinate variables, one can replace $ax+by+cz+dw$ with $z$; therefore, we learn that any bisection geometries locus in the four-section geometry (\ref{general complete intersection in 3.1}) admits a transformation to the complete intersection of the following form:
\begin{eqnarray}
\label{bisection complete intersection in 3.1}
2a_5\,xy+ z\, (2a_6\,x+2a_8\,y+a_3\,z+2a_{10}\,w) & =0 \\ \nonumber
b_1\, x^2+b_2\, y^2+b_3\, z^2+b_4\, w^2 + &  \\ \nonumber
2b_5\,xy+2b_6\,xz+2b_7\,xw+2b_8\,yz+2b_9\,yw+2b_{10}\,zw & = 0
\end{eqnarray}
$\{x=0, \hspace{1.5mm} z=0 \}$ and $\{y=0, \hspace{1.5mm} z=0 \}$ yield bisections.  
\par Now, we compute the associated double cover of the bisection geometries locus (\ref{bisection complete intersection in 3.1}). One subtracts $\lambda$ times the first equation from the second equation, and one arranges the coefficients of the resulting equation into a 4 $\times$ 4 symmetric matrix. ($\lambda$ serves as a variable.) Taking the double cover of the determinant of this symmetric matrix, one arrives at the equation of the associated double cover \cite{BM, Kimura1905}. The resulting associated double cover is given by
\begin{equation}
\label{double cover of bisection locus in 3.1}
\tau^2 = \begin{vmatrix}
b_1 & b_5-\lambda a_5 & b_6-\lambda  a_6 & b_7 \\
b_5-\lambda a_5 & b_2 & b_8-\lambda a_8 & b_9 \\
b_6-\lambda  a_6 & b_8-\lambda a_8 & b_3-\lambda a_3 & b_{10}-\lambda a_{10} \\
b_7 & b_9 & b_{10}-\lambda a_{10} & b_4
\end{vmatrix}.
\end{equation}
$| \cdot |$ on the right-hand side means to take the determinant of the matrix. Expanding the determinant on the right-hand side, we find that the term $e_0$ of $\tau^2 = e_0\lambda^4 + e_1\lambda^3 +e_2\lambda^2 + e_3\lambda +e_4$ is a perfect square: $e_0 = a_5^2\, a_{10}^2$. Thus, the double cover (\ref{double cover of bisection locus in 3.1}) actually has two global sections \cite{MTsection}, and therefore yields the Jacobian of the complete intersection (\ref{bisection complete intersection in 3.1}). Because the Jacobian (\ref{double cover of bisection locus in 3.1}) has two global sections, the Mordell--Weil rank is one, and we deduce that $U(1)\times \Z_2$ forms in F-theory on the bisection geometries locus (\ref{bisection complete intersection in 3.1}).
\par Because we demonstrated that every bisection geometries locus in the four-section geometry (\ref{general complete intersection in 3.1}) admits a transformation \footnote{Under transformation $x\rightarrow x+i\, y$ and $y\rightarrow x-i\, y$, one can confirm that the bisection geometries loci considered in \cite{Kdisc, Kimura1905} admit transformation to the complete intersections of the form (\ref{bisection complete intersection in 3.1}).\label{footnote12}} to the complete intersection (\ref{bisection complete intersection in 3.1}), we learn from these computations that $U(1)\times \Z_2$ forms in F-theory on any bisection geometries locus in the four-section geometry (\ref{general complete intersection in 3.1}).

\subsection{Enhancement of $U(1)$ to $SU(4)\times SU(2)\times SU(2)\times SU(2)$}
\label{section3.2}
\par Now, we would like to demonstrate that when some of the coefficients of the bisection geometry locus are set to zero, the $U(1)$ gauge group is enhanced to $SU(4)\times SU(2)\times SU(2)\times SU(2)$. We begin with the complete intersections of two quadrics, given by:
\begin{eqnarray}
\label{transformed complete intersection in 3.2}
a_1\, (x^2+z^2) + y\, (2a_5\,x+a_2\,y+2a_8\,z+2a_9\,w) & =0 \\ \nonumber
b_1\, x^2+b_2\, y^2+b_3\, z^2+b_4\, w^2 &  \\ \nonumber
+ 2b_5\,xy+2b_6\,xz+2b_7\,xw+2b_8\,yz+2b_9\,yw+2b_{10}\,zw & = 0
\end{eqnarray}
The four-section splits into a pair of bisections in these complete intersections. By an argument similar to that we noted in the footnote \ref{footnote12}, under the transformation $x\rightarrow x+z$, $z\rightarrow i\, (x-z)$ (followed by an exchange of the coordinates $(y,z)\rightarrow (z,y)$), the complete intersections (\ref{transformed complete intersection in 3.2}) transform to those in the bisection geometry locus (\ref{bisection complete intersection in 3.1}). $\{x+i\, z=0, \hspace{2mm} y=0\}$ and $\{x-i\, z=0, \hspace{2mm} y=0\}$ yield bisections. $U(1)\times \Z_2$ forms in 6D F-theory on the complete intersections (\ref{transformed complete intersection in 3.2}).
\par Next, we consider the situation where the coefficients $a_2, a_5, a_8$, $b_1, b_3, b_5, b_7, b_8, b_9, b_{10}$ are set to zero. That is, we consider the complete intersection of the following form:
\begin{eqnarray}
a_1\, (x^2+z^2) + 2a_9\,yw & =0 \\ \nonumber
b_2\, y^2+b_4\, w^2 +2b_6\,xz & = 0.
\end{eqnarray}
For the sake of notation clarity, we rename the coefficients $a_1, a_9, b_2, b_4, b_6$ as $f_1, f_2, h_1, h_2, h_3$, respectively, as follows:
\begin{eqnarray}
\label{SU(4) complete intersection in 3.2}
f_1\, (x^2+z^2) + 2f_2\,yw & =0 \\ \nonumber
h_1\, y^2+h_2\, w^2 +2h_3\,xz & = 0.
\end{eqnarray}
As we will see shortly, $SU(4)\times SU(2)\times SU(2)\times SU(2)\times \Z_2$ forms in 6D F-theory on the complete intersection (\ref{SU(4) complete intersection in 3.2}). 
\par The associated double cover of the complete intersection (\ref{SU(4) complete intersection in 3.2}) is given as follows:
\begin{eqnarray}
\label{SU(4) double cover in 3.2}
\tau^2 & = \begin{vmatrix}
-\lambda f_1 & 0 & h_3 & 0 \\
0 & h_1 & 0 & -\lambda f_2 \\
h_3 & 0 & -\lambda f_1 & 0 \\
0 & -\lambda f_2 & 0 & h_2
\end{vmatrix} \\ \nonumber
& = -f_1^2f_2^2\, \lambda^4+ (f_1^2h_1 h_2+ f_2^2 h_3^2) \lambda^2 - h_1h_2h_3^2.
\end{eqnarray}
The discriminant is given by 
\begin{equation}
\label{SU(4) discriminant in 3.2}
\Delta \sim f_1^2 f_2^2 h_1 h_2 h_3^2 \, (f_1^2 h_1h_2 - f_2^2 h_3^2)^4.
\end{equation}
\par Because the coefficient $e_0$ of the term $\lambda^4$ is a perfect square, $e_0=-f_1^2f_2^2$, the obtained associated double cover, has a global section \cite{MTsection}; thus, the double cover (\ref{SU(4) double cover in 3.2}) yields the Jacobian fibration of the complete intersection (\ref{SU(4) complete intersection in 3.2}). When the coefficient $e_1$ of the term $\lambda^3$ is nonzero, the double cover has two global sections and $U(1)$ forms in F-theory on the double cover \cite{MTsection}. For our case (\ref{SU(4) double cover in 3.2}), the coefficient $e_1$ is zero, $e_1=0$, and $U(1)$ is enhanced to $SU(2)$, as discussed in \cite{Kimura1907}.
\par The associated double cover (\ref{SU(4) double cover in 3.2}) admits a transformation to the following Weierstrass equation:
\begin{equation}
\label{SU(4) Weierstrass}
\begin{split}
y^2= & x^3+ (-\frac{1}{3}f_2^4h_3^4-\frac{14}{3}f_1^2f_2^2h_1h_2h_3^2-\frac{1}{3}f_1^4h_1^2h_2^2)\, x \\
& + (-\frac{2}{27}f_1^6h_1^3h_2^3+\frac{22}{9}f_1^4f_2^2h_1^2h_2^2h_3^2+\frac{22}{9}f_1^2f_2^4h_1h_2h_3^4-\frac{2}{27}f_2^6h_3^6). 
\end{split}
\end{equation}
From the discriminant (\ref{SU(4) discriminant in 3.2}), one can find that the 7-branes are wrapped on the following curves:
\begin{eqnarray}
\label{discriminant curves in 3.2}
C_1 & = \{f_1^2h_1h_2-f_2^2h_3^2=0\} \\ \nonumber
C_2 & =\{f_1=0\} \\ \nonumber
C_3 & =\{f_2=0\} \\ \nonumber
C_4 & =\{h_3=0\} \\ \nonumber
C_5 & =\{h_1=0\} \\ \nonumber
C_6 & =\{h_2=0\}.
\end{eqnarray}
From the equations (\ref{SU(4) discriminant in 3.2}) and (\ref{SU(4) Weierstrass}), it is clear that the singular fibers over the curve $C_1$ are type $I_4$, and the singular fibers over the curves $C_2$, $C_3$ and $C_4$ are type $I_2$. The fiber type over the curves $C_5$ and $C_6$ is $I_1$. 
\par Using an argument similar to that given in A.1 in \cite{Kdisc}, it is clear that the type $I_4$ fibers over the curve $C_1$ are split \cite{BIKMSV}. The $SU(4)$ gauge group forms on the 7-branes wrapped on the curve $C_1$, and the $SU(2)$ gauge group is supported on the three curves $C_2, C_3, C_4$. Therefore, the $SU(4)\times SU(2)\times SU(2)\times SU(2)\times \Z_2$ gauge group forms in F-theory on the complete intersection (\ref{SU(4) complete intersection in 3.2}).

\subsection{Transition from $U(1)\times \Z_2$ theory to $\Z_4$ via Higgsing}
\label{section3.3}
We demonstrate that the $SU(4)\times \Z_2$ gauge group in F-theory on the complete intersection (\ref{SU(4) complete intersection in 3.2}) transitions to $U(1)\times \Z_2$, and further transitions to a discrete $\Z_4$ gauge group through Higgsing processes. We deduce the matter spectrum in 6D F-theory on the complete intersection (\ref{SU(4) complete intersection in 3.2}) and show that this transition of the gauge group indeed occurs by giving vevs to hypermultiplets. We focus on the situations where the base surface is isomorphic to $\P^1\times \P^1$ and $\P^2$. (The results obtained in Section \ref{section3.1} and \ref{section3.2} hold for any base surface, when the degrees of the line bundles, sections of which yielding the coefficients of the complete intersections, are appropriately chosen to satisfy the Calabi--Yau condition \cite{Kimura1905}.)

\subsubsection{Case base is isomorphic to $\P^1\times\P^1$}
\label{section3.3.1}
We choose $f_1, f_2, h_1, h_2, h_3$ to be polynomials of bidegree (1,1) on $\P^1\times\P^1$. This amounts to regarding the equation (\ref{SU(4) complete intersection in 3.2}) as a (2,1,1) and (2,1,1) complete intersection in the product $\P^3\times\P^1\times\P^1$, yielding a Calabi--Yau 3-fold. The natural projection onto $\P^1\times \P^1$ yields a genus-one fibration. 
\par Because a bidegree $(a,b)$ curve has genus $(a-1)(b-1)$, the curve $C_1$ has genus $g=(4-1)(4-1)=9$. The other curves $C_2, C_3, C_4, C_5, C_6$ have genus 0. There should exist matter fields arising on the curve $C_1$ contributing to the genus $g=9$ because of the anomaly equations \cite{KPT1011}. We expect that nine adjoint hypermultiplets ${\bf 15}$ arise on the curve $C_1$. Similar to that discussed in \cite{BIKMSV}, considering further compactification on $T^2$ down to 4D theory, this agrees with the result obtained in \cite{KMP1996}. 
\par Because the curves $C_2, C_3, C_4, C_5, C_6$ have genus 0, only the fundamental representation ${\bf 2}$ (or ${\bf 1}$) can arise \cite{KPT1011, MTsection} on these curves.
\par By studying the discriminant (\ref{SU(4) discriminant in 3.2}), it is clear that the curves supporting the $SU(4)$ or $SU(2)$ gauge group intersect along: 
\begin{eqnarray}
\label{intersections of curves in 3.3.1}
\{f_1=0\} \cap \{f_2=0\}, \hspace{1.5mm} \{f_1=0\} \cap \{h_3=0\}, \hspace{1.5mm} \{h_1=0\} \cap \{f_2=0\},\\ \nonumber
\{h_1=0\} \cap \{h_3=0\}, \hspace{1.5mm} \{h_2=0\} \cap \{f_2=0\}, \hspace{1.5mm} \{h_2=0\} \cap \{h_3=0\}, \\ \nonumber
\{f_2=0\} \cap \{h_3=0\}.
\end{eqnarray}
Matter fields are localized at these intersections. Because two bidegree (1,1) curves intersect at two points, each of the seven intersections in (\ref{intersections of curves in 3.3.1}) consists of two points. The curve $C_1$ supporting $SU(4)$ and another curve supporting $SU(2)$ intersect for the first six intersections in (\ref{intersections of curves in 3.3.1}). The last intersection in (\ref{intersections of curves in 3.3.1}) is the intersection of the curves $C_3$ and $C_4$ at two points. A bifundamental $({\bf 2}, {\bf 2})$ arises at each of the two intersection points of $C_3$ and $C_4$.
\par A symmetry argument suggests that the identical matter representations arise from the four points in the intersections $\{f_1=0\} \cap \{f_2=0\}$, $\{f_1=0\} \cap \{h_3=0\}$; similarly, the identical matter arises from the eight points in the intersections $\{h_1=0\} \cap \{f_2=0\}$, $\{h_1=0\} \cap \{h_3=0\}$, $\{h_2=0\} \cap \{f_2=0\}$, $\{h_2=0\} \cap \{h_3=0\}$.
\par Because the base is isomorphic to $\P^1\times\P^1$, the number of tensor multiplets arising in 6D F-theory is $T=1$, and $H-V=273-29=244$ by the 6D anomaly cancellation condition \cite{GSW6d, Sagnotti, Erler, Sch6d, GM0005, KMT1008}. The non-Abelian gauge group forming in 6D F-theory compactification is $SU(4)\times SU(2) \times SU(2)\times SU(2)$; thus, $V=24$. Therefore, the number of hypermultiplets should be $H=268$ to cancel the anomaly. There are nine adjoint hypermultiplets ${\bf 15}$ of $SU(4)$ on the curve $C_1$, and two bifundamentals $({\bf 2}, {\bf 2})$ localized at the intersections of $C_3$ and $C_4$. To cancel the anomaly, it appears a unique choice that the matter representations have 12 dimensions at each of the four points at the intersections $\{f_1=0\} \cap \{f_2=0\}$, $\{f_1=0\} \cap \{h_3=0\}$, and have 8 dimensions at each of the eight points at the intersections $\{h_1=0\} \cap \{f_2=0\}$, $\{h_1=0\} \cap \{h_3=0\}$, $\{h_2=0\} \cap \{f_2=0\}$, $\{h_2=0\} \cap \{h_3=0\}$. Matter at each of the four points at the intersections $\{f_1=0\} \cap \{f_2=0\}$, $\{f_1=0\} \cap \{h_3=0\}$ is either ${\bf 6}\oplus {\bf 6}$ or $({\bf 6}, {\bf 2})$ \footnote{The possibility that the 12-dimensional matter representation is ${\bf 10}\oplus {\bf 2}$ is ruled out because of the anomaly cancellation condition. Matter ${\bf 10}$ of $SU(4)$ contributes to the genus \cite{KPT1011} of the curve $C_1$ if it arises on the curve; however, there are nine adjoint hypermultiplets ${\bf 15}$ on the curve $C_1$ as we mentioned previously. Therefore, the presence of ${\bf 10}$ violates the anomaly cancellation conditions.}, and matter at each of the eight points in the intersections $\{h_1=0\} \cap \{f_2=0\}$, $\{h_1=0\} \cap \{h_3=0\}$, $\{h_2=0\} \cap \{f_2=0\}$, $\{h_2=0\} \cap \{h_3=0\}$ is either $({\bf 4}, {\bf 2})$ or ${\bf 4}\oplus {\bf 4}$. Including the neutral hypermultiplets, one can confirm that the anomaly cancels with these matter. 
\par Turning on vevs for hypermultiplets breaks $SU(4)\times \Z_2$ gauge group down to $U(1)\times \Z_2$. Adjoint hypermultiplets ${\bf 15}$ of $SU(4)$ are used for Higgsing $SU(4)$ gauge group down to $U(1)$, and the remaining adjoint hypermultiplets become scalar fields of $U(1)$ charge 4, along the line of arguments as in \cite{MTsection}. One of the resulting charge 4 scalar fields can be used to Higgs the $U(1)$ further down to a discrete $\Z_4$ gauge group, and this process corresponds to deforming the coefficients of complete intersection, so after the deformation, the resulting complete intersection is away from the bisection geometry locus.

\subsubsection{Case base is isomorphic to $\P^2$}
We choose $f_1, f_2$ to be polynomials of degree two on $\P^2$, and $h_1, h_2, h_3$ to be polynomials of degree one on $\P^2$ here. The equation (\ref{SU(4) complete intersection in 3.2}) is (2,2) and (2,1) complete intersection in $\P^3\times \P^2$ for this situation, yielding a Calabi--Yau 3-fold. The natural projection onto $\P^2$ yields a genus-one fibration.
\par In this situation, because the curve $C_1$ has degree 6, $C_1$ has genus $g=\frac{1}{2}(6-1)(6-2)=10$. We expect that there are ten adjoint hypermultiplets of $SU(4)$ on the curve $C_1$. The other curves, $C_2, C_3, C_4, C_5, C_6$, have genus 0. The two curves $C_3$ and $C_4$ intersect at two points, and a bifundamental $({\bf 2}, {\bf 2})$ arises at each of the two points.
\par Matter fields are localized at the intersections of the curves (\ref{intersections of curves in 3.3.1}). $C_1$ supporting $SU(4)$ intersects with other curves at the first six intersections in (\ref{intersections of curves in 3.3.1}), and the last intersection in (\ref{intersections of curves in 3.3.1}) is the intersection of the two curves $C_3$ and $C_4$.
\par Because the base is isomorphic to $\P^2$, the number of tensor multiplets arising in 6D F-theory is $T=0$. The 6D anomaly cancellation condition is then $H-V=273$; therefore, the anomaly cancellation condition reads $H=297$. Utilizing an argument similar to that given in Section \ref{section3.3.1}, it is clear that matter representations arising from the six points at the intersections $\{f_1=0\} \cap \{f_2=0\}$, $\{f_1=0\} \cap \{h_3=0\}$ should be identical. Similarly, the identical matter arises from the six points at the intersections $\{h_1=0\} \cap \{f_2=0\}$, $\{h_1=0\} \cap \{h_3=0\}$, $\{h_2=0\} \cap \{f_2=0\}$, $\{h_2=0\} \cap \{h_3=0\}$. When the matter representation of 12 dimensions arise from the six points in $\{f_1=0\} \cap \{f_2=0\}$, $\{f_1=0\} \cap \{h_3=0\}$ and matter of 8 dimensions arise from the six points in $\{h_1=0\} \cap \{f_2=0\}$, $\{h_1=0\} \cap \{h_3=0\}$, $\{h_2=0\} \cap \{f_2=0\}$, $\{h_2=0\} \cap \{h_3=0\}$, the anomaly cancels, including the neutral hypermultiplets. The dimensions of matter representations arising from the twelve intersection points appear to be unique. Matter localized at each of the six points at the intersections $\{f_1=0\} \cap \{f_2=0\}$, $\{f_1=0\} \cap \{h_3=0\}$ is either ${\bf 6}\oplus {\bf 6}$ or $({\bf 6}, {\bf 2})$, and matter localized at each of the six points at the intersections $\{h_1=0\} \cap \{f_2=0\}$, $\{h_1=0\} \cap \{h_3=0\}$, $\{h_2=0\} \cap \{f_2=0\}$, $\{h_2=0\} \cap \{h_3=0\}$ is either $({\bf 4}, {\bf 2})$ or ${\bf 4}\oplus {\bf 4}$.
\par The $SU(4)\times \Z_2$ gauge group is Higgsed to $U(1)\times \Z_2$ by turning on vevs for hypermultiplets, and adjoint hypermultiplets ${\bf 15}$ of $SU(4)$ are used to break the gauge group in this process. The remaining adjoint hypermultiplets ${\bf 15}$ become scalar fields of $U(1)$ charge 4, one of which can be used to Higgs the $U(1)$ down to a discrete $\Z_4$ gauge group.

\section{Construction of models with $U(1)\times \Z_4$}
\label{section4}
If we can construct complete intersections of two quadrics in $\P^3$ fibered over a base whose Jacobians have the Mordell--Weil rank one, this yields a family of Calabi--Yau genus-one fibrations on which $U(1)\times \Z_4$ forms in F-theory, and it suffices to check either the coefficient $e_0$ or $e_4$ of the associated double cover is a perfect square to achieve this construction. We present an explicit example of such a family here. 
\par The coefficients $e_0$ and $e_4$ of the associated double cover of complete intersection of the general form (\ref{general complete intersection in 3.1}) are not perfect squares, so some specific coefficients of the complete intersection need to be chosen. 
\par To this end, we choose the following specific complete intersection:
\begin{eqnarray}
\label{specific complete intersection in 4}
a_1\, x^2+a_2\, y^2+a_3\, z^2+a_4\, w^2 +2a_6\,xz+2a_9\,yw & =0 \\ \nonumber
b_1\, x^2+b_2\, y^2+b_1\, z^2+b_2\, w^2 +2b_2\,xz+2b_1\,yw & =0
\end{eqnarray}
This is a four-section geometry, and as we demonstrate shortly, the associated double cover of this complete intersection has two global sections. 
\par The associated double cover is given by the following equation:
\begin{equation}
\label{double cover in 4}
\tau^2 = \begin{vmatrix}
b_1-\lambda  a_1 & 0 & b_2-\lambda  a_6 & 0 \\
0 & b_2-\lambda  a_2 & 0 & b_1-\lambda a_9 \\
b_2-\lambda  a_6 & 0 & b_1-\lambda a_3 & 0 \\
0 & b_1-\lambda a_9 & 0 & b_2-\lambda  a_4
\end{vmatrix}.
\end{equation}
Expanding the determinant on the right-hand side, we find that the term $e_4$ of $\tau^2 =e_0\lambda^4 + e_1\lambda^3 +e_2\lambda^2 + e_3\lambda +e_4$ is a perfect square, $e_4 = - (b_1^2-b_2^2)^2$; therefore, the associated double cover (\ref{double cover in 4}) has two global sections. Thus, the associated double cover (\ref{double cover in 4}) yields the Jacobian fibration of the complete intersection (\ref{specific complete intersection in 4}), and the Mordell--Weil rank of the Jacobian is one. $U(1)\times \Z_4$ forms in F-theory on the complete intersections (\ref{specific complete intersection in 4}).

\section{Concluding remarks and open problems}
\label{section5}
\par We observed in Section \ref{section3.1} an expansion of a discrete gauge group along any bisection geometry locus in four-section geometry, realized as complete intersections of two quadric hypersurfaces in $\P^3$ fibered over any base. $U(1)$ in $U(1)\times \Z_2$ forming along the bisection geometry locus in the four-section geometry is enhanced to $SU(4)\times SU(2)\times SU(2)\times SU(2)$ when some of the coefficients of the complete intersection are set to zero, as discussed in Section \ref{section3.2}. We deduced matter spectra on these enhanced 6D F-theory models. Starting from these enhanced 6D models, by giving vevs to hypermultiplets, we demonstrated that $U(1)\times \Z_2$ theory indeed transitions to a model with a discrete $\Z_4$ gauge group through Higgsing, when the base surfaces are isomorphic to $\P^1\times \P^1$ and $\P^2$. This observation can support, at least to some level, the possible interpretation proposed in \cite{Kimura1907} that can resolve the puzzle raised in \cite{Kimura1905}. Extending the analysis in Section \ref{section3.3} to 6D F-theory models over other base surfaces is a likely direction for future study.
\par It might be interesting to determine if the family (\ref{specific complete intersection in 4}) on which $U(1)\times \Z_4$ forms in F-theory constructed in Section \ref{section4} belongs to some four-section geometry loci of a multisection geometry of higher degree. If this is the case, this would also support the proposal in \cite{Kimura1907}. Analyzing the Higgsing processes occurring in this geometry is a likely direction for future study.

\section*{Acknowledgments}

We would like to thank Shun'ya Mizoguchi and Shigeru Mukai for discussions.

\end{document}